\documentclass[11pt,preprint]{aastex}
\newcommand{\etal}{{\it et~al.}}

\begin{document}

\title{Main Belt Asteroids with WISE/NEOWISE: Near-Infrared Albedos}

\author{Joseph R. Masiero\altaffilmark{1}, T. Grav\altaffilmark{2}, A. K. Mainzer\altaffilmark{1}, C. R. Nugent\altaffilmark{1}, J. M. Bauer\altaffilmark{1,3},  R. Stevenson\altaffilmark{1}, S. Sonnett\altaffilmark{1}}

\altaffiltext{1}{Jet Propulsion Laboratory/Caltech, 4800 Oak Grove Dr., MS 183-601, Pasadena, CA 91109, {\it Joseph.Masiero@jpl.nasa.gov; amainzer@jpl.nasa.gov; cnugent@jpl.nasa.gov; James.Bauer@jpl.nasa.gov; Rachel.A.Stevenson@jpl.nasa.gov; sarah.sonnett@jpl.nasa.gov}}
\altaffiltext{2}{Planetary Science Institute, Tucson, AZ {\it tgrav@psi.edu}}
\altaffiltext{3}{Infrared Processing and Analysis Center, Caltech, Pasadena, CA}

\begin{abstract}
We present revised near-infrared albedo fits of $2835$ Main Belt
asteroids observed by WISE/NEOWISE over the course of its fully
cryogenic survey in 2010.  These fits are derived from reflected-light
near-infrared images taken simultaneously with thermal emission
measurements, allowing for more accurate measurements of the
near-infrared albedos than is possible for visible albedo
measurements.  As our sample requires reflected light measurements, it
undersamples small, low albedo asteroids, as well as those with blue
spectral slopes across the wavelengths investigated.  We find that the
Main Belt separates into three distinct groups of $6\%$, $16\%$, and
$40\%$ reflectance at $3.4~\mu$m.  Conversely, the $4.6~\mu$m albedo
distribution spans the full range of possible values with no clear
grouping.  Asteroid families show a narrow distribution of $3.4~\mu$m
albedos within each family that map to one of the three observed
groupings, with the (221) Eos family being the sole family associated
with the $16\%$ reflectance $3.4~\mu$m albedo group.  We show that
near-infrared albedos derived from simultaneous thermal emission and
reflected light measurements are an important indicator of asteroid
taxonomy and can identify interesting targets for spectroscopic
followup.

\end{abstract}

\section{Introduction}
The Wide-field Infrared Survey Explorer \citep[WISE,][]{wright10}
performed an all-sky survey in the thermal infrared, imaging each
field of view simultaneously in four infrared wavelengths during the
fully cryogenic portion of the mission and in the two shortest
wavelengths when the mission continued as the Near-Earth Object WISE
survey \citep[NEOWISE,][]{mainzer11nw}.  The four WISE bandpasses are
referred to as W1, W2, W3, and W4, and cover the wavelength ranges of
$3.1-3.8~\mu$m, $4.1-5.2~\mu$m, $7.6-16.3~\mu$m, and $19.8-23.4~\mu$m
respectively, with photometric central wavelengths of $3.4~\mu$m,
$4.6~\mu$m, $12~\mu$m, and $22~\mu$m respectively \citep{wright10}.

The single-frame WISE/NEOWISE data allow us to investigate the thermal
emission and reflectance properties of the minor planets of the Solar
system.  Due to their proximity to the Sun, near-Earth objects
(objects with perihelia $q<1.3~$AU) are typically dominated by thermal
emission in W2 (and occasionally even in W1), while for more distant
objects W2 measures a combination of reflected and emitted light and
W1 is dominated by reflected light.  For all minor planets observed by
NEOWISE, bands W3 and W4 are dominated by thermal emission.

Measurements of the thermal emission from asteroids can be used to
determine the diameter of these bodies through the application of
thermal models such as the Near Earth Asteroid Thermal Model
\citep[NEATM,][]{harrisNEATM}.  NEATM assumes that the asteroid is a
non-rotating sphere with no emission from the night side, and a
variable beaming parameter ($\eta$) is used to account for variability
in thermophysical properties and phase effects.  NEATM provides a
rapid method of determining diameter from thermal emission data that
is reliable to $\sim10\%$ when the beaming parameter can be fit
\citep{mainzer11cal}.  Visible light measurements available from the
Minor Planet Center (MPC)\footnote{http://www.minorplanetcenter.net}
can then be combined with these models to constrain the geometric
albedo at visible wavelengths ($p_V$).  However, as these data are not
simultaneous with the thermal infrared measurements, uncertainties due
to rotation phase, observing geometry, and photometric phase behavior
instill significant systematic errors in $p_V$ determinations.  The
preliminary asteroid thermal fits presented in
\citet{mainzer11neo,mainzer12pc,masiero11,masiero12pc,grav11troj,grav11hilda}
for the near-Earth objects, Main Belt asteroids, Hildas, and Jupiter
Trojans account for these uncertainties in the determination of the
optical $H$ magnitude resulting in a larger relative error on albedo
than is found for diameter.

For objects that were observed by NEOWISE in both thermal emission and
near-infrared (NIR) reflected light, we can simultaneously constrain
the diameter as well as the NIR albedo.  As these data were taken at the
same time and observing conditions as the thermal data used to model
the diameter, no assumptions are needed regarding the photometric
phase behavior of these objects, and light curve changes from rotation
or viewing geometry do not contribute to the uncertainty.  These NIR
albedos will thus be a more precise indicator of the surface
properties than the visible albedos.  

The behavior of the NIR region of an asteroid's reflectance spectrum
can be used as a probe of the composition of the surface.  Spectra of
asteroids in the NIR have been used for taxonomic classification
\citep{demeo09}, to constrain surface mineralogy
\citep{gaffey02,reddy12vesta}, and to search for water in the Solar
system \citep{rivkin10,campins10}.  Near-infrared albedo measurements
have also been used to identify candidate metal-rich objects in the NEO
population \citep{harris14}.  \citet{mainzer11tax} use the ratio of
the NIR and visible albedos as a proxy for spectral slope, and show a
correspondence between this ratio and various taxonomic
classifications.  This relation was used by \citet{mainzer11neo} to
determine preliminary classifications for NEOs, while
\citet{grav11hilda} and \citet{grav12tax} expanded upon this technique
to taxonomically classify Hilda and Jupiter Trojan asteroids.

\citet{masiero11} presented NIR albedo measurements of Main Belt
asteroids assuming that the albedos at the W1 and W2 wavelengths were
identical.  However, for objects that have a sufficient number of
detections of reflected light in multiple NIR bands we can
independently constrain each albedo ($p_{W1}$ and $p_{W2}$ for the W1
and W2 bandpasses respectively).  In this work, we present new thermal
model fits of the NIR albedos of Main Belt asteroids (MBAs), allowing
$p_{W1}$ and $p_{W2}$ to vary independently.  These albedos allow us
to better distinguish different MBA compositional classes.  They are
also particularly useful for investigations of collisional families
seen in the Main Belt, which show strongly correlated physical
properties within each family.

\section{Data and Revised Thermal Fits}
\label{sec.data}

To fit for NIR albedos of Main Belt asteroids, we use data from
WISE/NEOWISE all-sky single exposure source table, which are available
for download from the Infrared Science Archive
\citep[IRSA\footnote{http://irsa.ipac.caltech.edu}][]{cutriExpSupp}.
We extract photometric measurements of all asteroids observed by WISE
following the technique described in \citet{masiero11} and
\citet{mainzer11nw}.  In particular, we use the NEOWISE observations
reported to the MPC and included in the MPC's minor planet observation
database as the final validated list of reliable NEOWISE detections of
Solar system objects.

For objects with WISE detections in all four bands, we follow the
fitting technique described in \citet{grav12tax} to independently
determine the albedos in bands W1 and W2.  This technique uses a
faceted sphere with a temperature distribution drawn from the NEATM
model to calculate the predicted visible and infrared magnitudes for
each object.  Diameter, beaming parameter, and visible, W1, and W2
albedos are all varied until a best-fit is found.  Monte carlo
simulations of the data using the measurement errors then provide a
constraint on the uncertainty of each parameter.  We require at least
three detections in each band above SNR=4 to use that band in our fit.
Main Belt asteroids are typically closer to the Sun at the time of
observation than the Trojans and Hildas discussed in
\citet{grav12tax}, and thus the measured W2 flux can have a larger
contribution of thermal emission for MBAs.  Flux in the W1 band is
typically dominated by reflected light for MBAs observed by WISE,
although low albedo objects ($p_V<0.1$) at heliocentric distances of
$R_{\sun}<2.5~$AU can be thermally dominated in W1 as well.

In order to ensure reliable fits for W2 albedos ($p_{W2}$), we require
that the beaming parameter be fit by the model.  The beaming parameter
is a variable in the NEATM fit that consolidates uncertainties in the
model due to viewing geometries and surface thermophysical parameters,
and can be characterized as an enhancement of the thermal emission in
the direction of the Sun.  Changes in thermal properties or phase
angles will lead to a range of possible beaming parameters for
MBAs. In order to fit the beaming parameter, our model requires
detections of thermal emission in two bands.  We also require that the
fraction of flux in W2 from reflected light be at least $10\%$ of the
total flux measured to fit $p_{W2}$.  While this should be sufficient
to constrain the W2 albedo in most cases, uncertainty in the beaming
parameter can lead to large uncertainties in $p_{W2}$.  To fit W1
albedo ($p_{W1}$), we followed the same procedure for $p_{W2}$, except
now requiring the W1 reflected light to be at least $50\%$ of the
total flux.  All objects that fulfilled the above requirements had
optically measured magnitudes available in the literature, and thus
allowed us to fit a visible albedo as well.

We present our updated thermal model fits for all objects satisfying
the above constraints in Table~\ref{tab.fits}, where we give the
object's name in MPC-packed format, absolute $H_V$ magnitude and $G$
photometric slope parameter from the MPC orbit file, associated family
from \citet{masiero13} (or ``...'' if the object is not associated to
a family), and our best fit and associated uncertainty on diameter,
beaming parameter ($\eta$), $p_{W1}$, and $p_{W2}$ if the latter could
be constrained (``...'' otherwise).  Objects with two epochs of
coverage have each epoch listed separately.  All errors are
statistical and do not include the systematic errors of $\sim10\%$ on
diameter and $\sim20\%$ on visible albedo
\citep[cf.][]{mainzer11cal,mainzer11iras,masiero11}.  Systematic
errors will increase the absolute error on the fitted quantities, but
do not affect relative comparisons within our sample, which is the
main goal of this paper.  This table contains $3080$ fits of $2835$
unique objects: $709$ fits of $679$ unique objects with constrained
$p_{W1}$ and $p_{W2}$; $2371$ fits of $2219$ unique objects with only
$p_{W1}$ constrained; $63$ objects that had one epoch where both NIR
albedos were constrained and one epoch where only $p_{W1}$ could be
fit.

We note that some of the fits for diameter and beaming parameter (and
thus albedo) are different from those presented in \citet{masiero11}.
Fits from NEATM using the same data set will give different values for
the diameter as the beaming parameter is varied.  In this case, by
independently considering $p_{W1}$ and $p_{W2}$, as opposed to
averaging over both for a single value of $p_{IR}$, the calculated
contribution of thermal flux in W2 will vary, which will result in a
refined value of $\eta$ and therefore diameter.  For the majority of
cases, diameters are consistent to within $10\%$ of the previous
value, visible albedos are consistent within $20\%$, and infrared
albedo and beaming parameters are consistent within $15\%$.  Revised
beaming parameters tend to be $\sim5\%$ smaller, making diameters
$\sim3\%$ smaller and visible albedos $\sim6\%$ larger.  W1 infrared
albedos tend to increase $\sim8\%$, however we are not necessarily
comparing similar quantities as the previous fits assumed W1 and W2
albedos were equal, allowing W2 measurements to alter the best-fit
value.
 
\begin{table}[ht]
\begin{center}
\caption{Revised thermal fits of Main Belt asteroids.
  Table~\ref{tab.fits} is published in its entirety in the electronic
  edition; a portion is shown here for guidance regarding its form and
  content}
\vspace{1ex}
{\tiny
\noindent
\begin{tabular}{ccccccccc}
\tableline
Name & diameter (km) & $\eta$ & $p_V$ & $p_{W1}$ & $p_{W2}$ & $H_V$ & G & Family\\
  \tableline
  00005 & 108.29$\pm$  3.70 & 0.87$\pm$0.10 & 0.27$\pm$0.03 & 0.37$\pm$0.03 & 0.20$\pm$0.10 &  6.9 & 0.15 & 00005\\
  00006 & 195.64$\pm$  5.44 & 0.91$\pm$0.09 & 0.24$\pm$0.04 & 0.35$\pm$0.03 & 0.48$\pm$0.32 &  5.7 & 0.24 &  ... \\
  00008 & 147.49$\pm$  1.03 & 0.81$\pm$0.01 & 0.23$\pm$0.04 & 0.38$\pm$0.04 &      ...      &  6.4 & 0.28 & 00008\\
  00009 & 183.01$\pm$  0.39 & 0.86$\pm$0.01 & 0.16$\pm$0.03 & 0.33$\pm$0.01 &      ...      &  6.3 & 0.17 &  ... \\
  00009 & 184.16$\pm$  0.90 & 0.78$\pm$0.01 & 0.16$\pm$0.01 & 0.34$\pm$0.04 &      ...      &  6.3 & 0.17 &  ... \\
  00011 & 142.89$\pm$  1.01 & 0.78$\pm$0.02 & 0.19$\pm$0.02 & 0.35$\pm$0.03 &      ...      &  6.6 & 0.15 &  ... \\
  00012 & 115.09$\pm$  1.20 & 0.84$\pm$0.02 & 0.16$\pm$0.03 & 0.32$\pm$0.02 &      ...      &  7.3 & 0.22 & 00012\\
  00014 & 140.76$\pm$  8.41 & 0.84$\pm$0.15 & 0.27$\pm$0.04 & 0.39$\pm$0.06 & 0.19$\pm$0.16 &  6.3 & 0.15 &  ... \\
  00015 & 231.69$\pm$  2.23 & 0.79$\pm$0.04 & 0.25$\pm$0.04 & 0.40$\pm$0.04 &      ...      &  5.3 & 0.23 & 00015\\
  00017 &  84.90$\pm$  2.03 & 0.77$\pm$0.04 & 0.19$\pm$0.03 & 0.37$\pm$0.08 &      ...      &  7.8 & 0.15 &  ... \\
\hline
\end{tabular}
}
\label{tab.fits}
\end{center}
\end{table}

The criteria we apply to our fits, discussed above, result in
selection effects in our sample that conspire to under-represent the
lowest albedo asteroids, as these objects are more likely to fall
below our detectability threshold.  This bias is a fundamental result
of requiring reflected light observations in W1 and/or W2, but does
not affect population surveys based on emission in W3 or W4.  Our
sample requirements also drive us to only use the data from the fully
cryogenic portion of the WISE survey.  It is possible to use fully
cryogenic data to constrain the diameter and beaming parameter, and
later measurements from either the NEOWISE post-cryo survey or the
recently restarted NEOWISE mission \citep{mainzer14restart} during a
brighter apparition to constrain the NIR albedo properties
\citep[cf.][]{grav12tax}, although this technique requires a different
method of handling that addresses difference in viewing geometry.  We
will apply this technique to Main Belt asteroids in future work.

For objects with a sufficient number of detections in W1 but below our
W2 sensitivity limit or our threshold for the fraction of reflected
light in W2, we determine only the W1 albedo.  These objects are
either too small to reflect a detectable amount of light in W2, may be
dominated by thermal emission in W2 (common for low albedo objects,
$p_{W1}<0.1$), or may have a blue spectral slope over the $3-5~\mu$m
range and thus ``drop out'' of W2.  Each of these scenarios will have
a different implication for interpreting the distribution of W2
albedos, most notably that our data are least sensitive to smaller,
lower albedo objects, as well as objects with blue $p_V$-$p_{W2}$ or
$p_{W1}$-$p_{W2}$ spectral slopes.  Interpretation of the distribution
of NIR albedos or spectral slopes, particularly as a function of
taxonomy or size, must thus be made with the appropriate caveats.  We
also explore stacking of the predicted positions of these object in W2
to recover drop-out objects in future work.

\section{Discussion}
\subsection{Albedo comparisons}

\begin{figure}[ht]
\begin{center}
\includegraphics[scale=0.4]{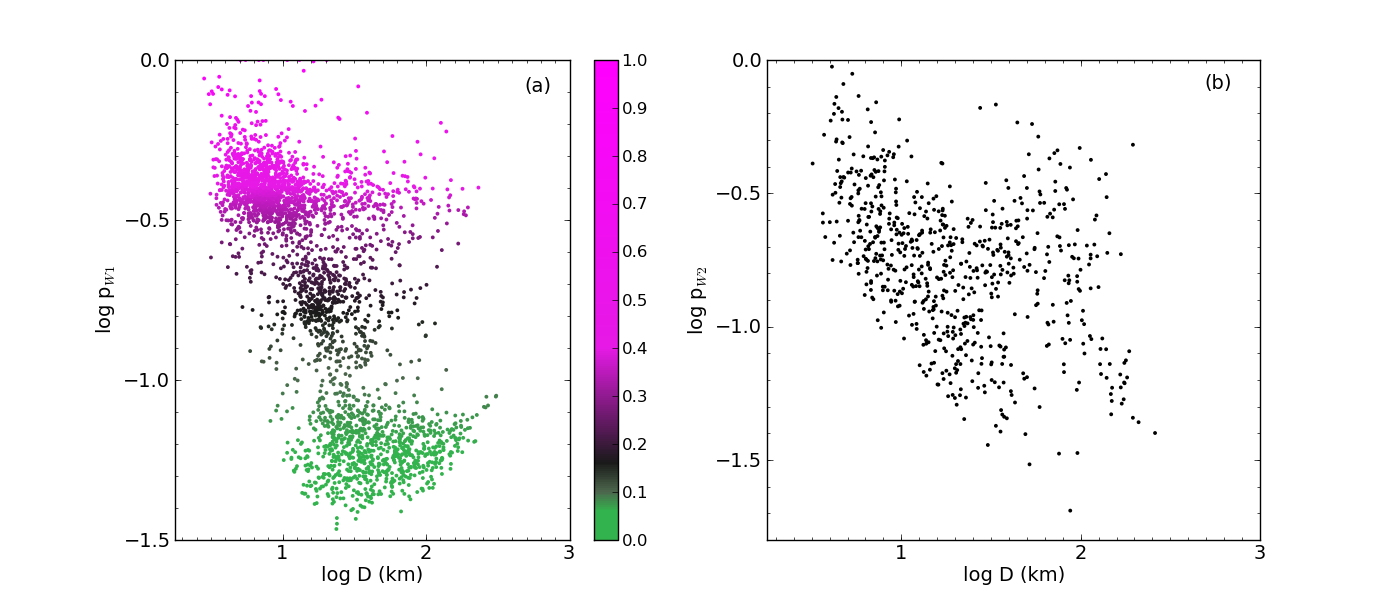}
\caption{(a) W1 infrared albedo ($p_{W1}$) compared to fitted
  diameter, where color also indicates $p_{W1}$ (as used in
  Figure~\ref{fig.eiFamBack}).  (b) W2 infrared albedo ($p_{W2}$)
  compared to diameter.  As the W1 and W2 detections are a measurement
  of reflected light they are strongly biased by albedo.  The dearth
  of small, low albedo objects in this plot that are observed in
  \citet{masiero11} is an artifact of this bias.}
\label{fig.w1alb}
\end{center}
\end{figure}

Figure \ref{fig.w1alb} shows $p_{W1}$ and $p_{W2}$ for all Main Belt
asteroids with sufficient data to constrain these parameters, compared
to the fitted diameter.  The W1 band is more sensitive than the W2
band \citep[single-frame $5\sigma$ sensitivity of $\sim0.22~$mJy vs
  $\sim0.31~$mJy respectively,][]{cutriExpSupp}, and W1 detections are
less frequently contaminated by thermal emission, meaning we are able
to measure $p_{W1}$ for more asteroids than have $p_{W2}$ measurements
($2835$ vs $679$).  Both data sets show a strong bias against small,
low-albedo asteroids, as is expected for data that require measurement
of a reflected light component.  A further bias against dark objects
in $p_{W2}$ due to rising thermal emission overtaking the small
reflected light component is also present.  From the data available we
see no evidence for a non-uniform distribution of $p_{W2}$, in
contrast to $p_{W1}$ which shows three significant albedo clumps at
$p_{W1}\sim0.06$, $p_{W1}\sim0.16$, and $p_{W1}\sim0.4$.

Visible albedos for over $136,000$ Main Belt asteroids were presented
in \citet{masiero11} and \citet{masiero12pc}.  These measurements were
based on the conversion of apparent visible magnitudes from a wide
range of predominantly ground-based surveys to absolute $H_V$
magnitudes when the orbit was determined by the MPC.  Absolute
magnitude is then converted to a predicted apparent magnitude
during the epoch of the WISE/NEOWISE observations, often after
assuming a photometric $G$ parameter \citep[cf.][]{bowell89} and
assuming rotational variations are averaged over during the set of
thermal infrared observations.  These conversions and assumptions will
instill additional uncertainty in the $p_V$ determinations beyond what
would be expected from uncertainties in the flux measurements and from
thermal modeling.  As a result, the fractional error on $p_V$ is
typically $50-100\%$ larger than the fractional error on diameter from
thermal model fits.

\begin{figure}[ht]
\begin{center}
\includegraphics[scale=0.6]{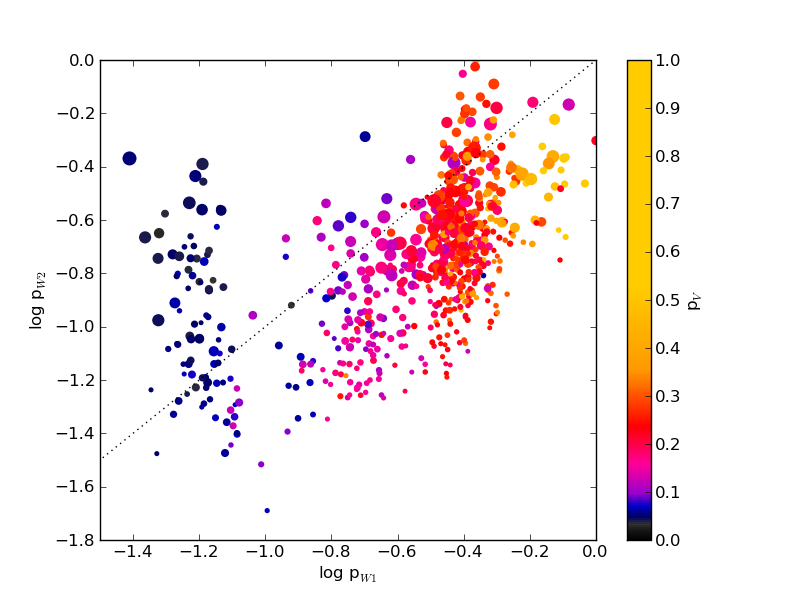}
\caption{W2 albedo vs W1 albedo for $679$ Main Belt asteroids.  The
  color of the points indicate their visible albedo following the
  colormap presented in \citet{masiero11} and shown in the colorbar,
  while the size of the point traces the fraction of the W2 flux that
  was due to reflected light.  The dotted line shows a 1-to-1
  correspondence; objects below the line will have a blue spectral
  slope from W1 to W2, while objects above the line will have a red
  slope.}
\label{fig.triAlb}
\end{center}
\end{figure}

\begin{figure}[ht]
\begin{center}
\includegraphics[scale=0.4]{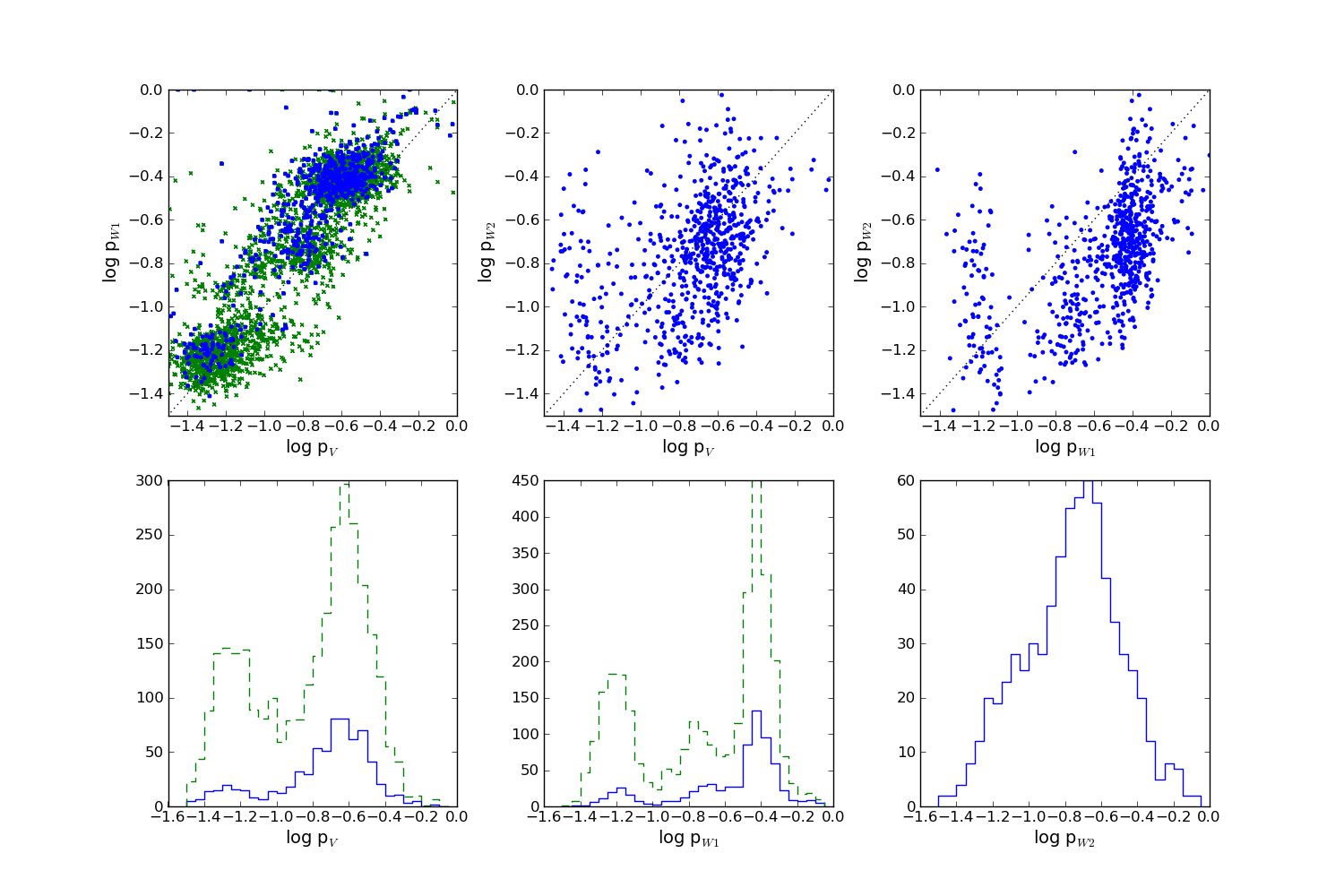}
\caption{Comparisons (top) and histograms (bottom) of asteroid albedos
  at visible ($p_V$), W1 ($p_{W1}$), and W2 ($p_{W2}$) wavelengths.
  Blue points/lines show objects with all three albedos fit by the
  thermal model ($679$ objects), while green crosses/dashed lines show
  objects with only visible and W1 fits ($2835$ objects).  Plots of
  $p_{W2}$ do not include green crosses as these objects do not have
  this parameter constrained by the model fits.  Dotted black lines in
  the comparison plots indicate a 1-to-1 relation.  While visible and
  W1 albedos show clear clumping, W2 albedos show no separation within
  our measurement errors.}
\label{fig.scatAlb}
\end{center}
\end{figure}

We compare the NIR albedos presented here to these visible albedos in
Figures~\ref{fig.triAlb} and \ref{fig.scatAlb}.  For the majority of
objects, $p_{W1}$ traces $p_V$ and can thus be used as an analog when
$p_V$ is not available.  The uncertainties of the $p_{W1}$
measurements in our data are smaller than the errors on $p_V$, and
thus act as a better constraint of the surface properties.  The
relationship between $p_{W2}$ and both $p_V$ and $p_{W1}$ is less
distinct, and varies over a large range of values for objects spanning
high and low $p_V$ and $p_{W1}$ albedos.

Comparing the $p_V$ distribution to Figure 10 from \citet{masiero11},
we see that our sample contains significantly fewer low albedo objects
than would be expected from a random sample of all Main Belt
asteroids.  The lack of low albedo objects is due primarily to the
observational selection effect imprinted on our dataset by the
requirement that the objects be detected in W1 and/or W2 in reflected
light.  This bias will increase as albedo decreases, preferentially
selecting objects with higher albedos.  A survey with deeper
sensitivity in these wavelengths would allow us to probe smaller sizes
at all albedos, but would still be subject to the same observational
biases.

\clearpage

\clearpage

Following \citet{grav12tax}, we can use our albedo measurements as a
proxy for spectral slope from visible wavelengths through the NIR.
Objects that fall above the 1-to-1 relationship in the top portion of
Figure~\ref{fig.scatAlb} will have a red spectral slope across the
wavelengths plotted, while objects below this relation will have a
blue spectral slope.  As $p_{W2}$ is the most poorly probed of the
three parameters, there will be inherent detection biases against blue
spectral slopes from objects that ``drop out'' and fall below our W2
detection threshold.  For this reason objects with the bluest slopes,
particularly low-albedo objects, will be under-represented in our fits
of $p_{W2}$.

\begin{figure}[ht]
\begin{center}
\includegraphics[scale=0.4]{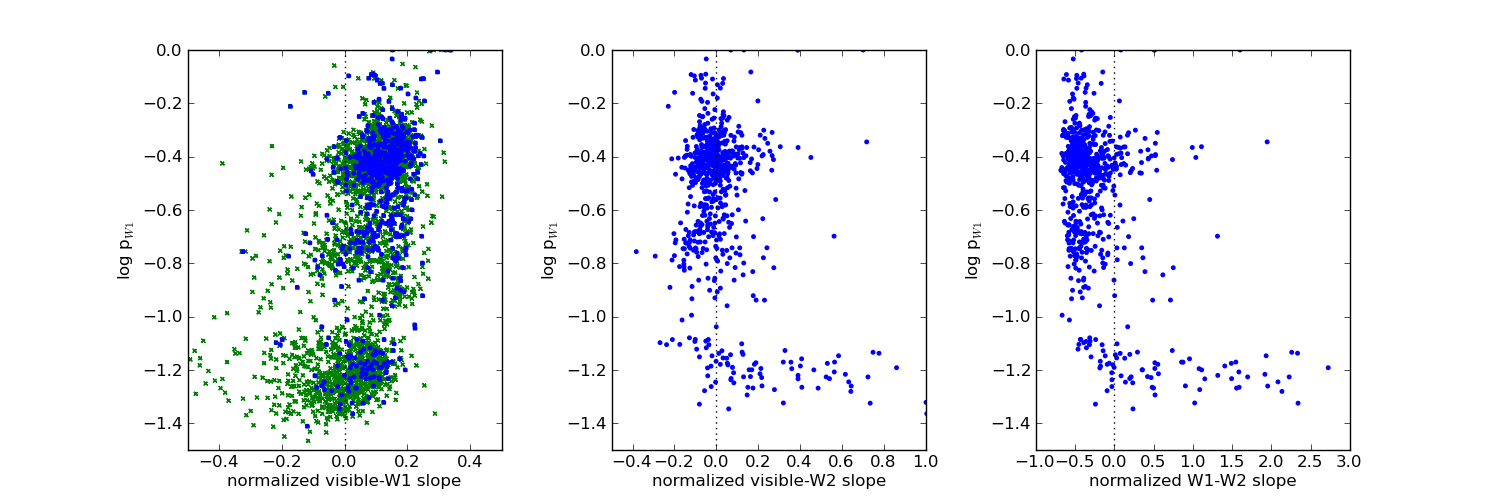}
\caption{Spectral slope, normalized to the W1 albedo, over
  visible-to-W1 wavelengths (left), visible-to-W2 wavelengths (center)
  and W1-to-W2 wavelengths (right), compared to the W1 albedo.  The
  dotted line shows a neutral slope; objects with positive slope
  values have red spectra and objects with negative values have blue
  spectra. High albedo objects tend to be red-sloped from
  visible to W1, and blue sloped from W1 to W2, while low albedo
  objects tend to be flat or red across the whole range.  Color/shape
  of points is the same as used in Figure~\ref{fig.scatAlb}.}
\label{fig.slopeComp}
\end{center}
\end{figure}

To better compare the spectral slope information, in
Figure~\ref{fig.slopeComp} we show the difference in albedo between
$p_V$, $p_{W1}$, and $p_{W2}$ normalized to the measured $p_{W1}$
value.  Objects with positive values have a red spectral slope, while
objects with negative values have a blue slope.  High albedo objects
($p_{W1}>0.1$) tend to show red slopes from visible to W1 wavelengths,
and then blue slopes between W1 and W2.  This behavior is similar to
what is observed for Eucrite meteorites at these wavelengths
\citep{reddy12hed}, and what would be expected from extrapolating a
typical S-type asteroid spectrum \citep{demeo09}.  High albedo objects
without a measured $p_{W2}$ albedo show similar visible-W1 slopes to
those with a measured $p_{W2}$.

Low albedo objects ($p_{W1}<0.1$) behave quite differently from their
high-albedo counterparts.  While slightly red from the visible to W1,
these objects show a wide range of visible-W2 and W1-W2 slopes, from
neutral in color to very red.  An important caveat to this is shown by
the objects without $p_{W2}$ fits, which have slopes ranging from
moderately red to significantly blue.  Blue-sloped objects would be
much fainter in W2 than W1 and thus would drop out from detection or
be dominated by thermal emission in W2.  It is probable that there is
a population of these objects with blue W1-W2 slopes that are not
represented in our plots.  Extrapolting from the NIR spectra of low
albedo objects from \citet{demeo09}, we associate our objects that
have red visible-W1 slopes with C-type and D-type objects.  We can
similarly associate the objects having blue-slopes with B-type
asteroids, however we note that only $\sim1\%$ of objects studied by
\citet{demeo09} were identified as B-type asteroids, while $\sim10\%$
of the asteroids in our study have low albedo and blue spectral slope.
From \citet{neesePDS} we find that the majority of our blue sloped
objects that have Bus-DeMeo taxonomic classifications are identified
as B or Ch class objects, the latter of which represents a fraction of
the spectroscopic sample comparable to the fraction of our sample in
this group.  Our blue slope may be indicative of the presence of
mineralogical absorption features in the spectra of low albedo objects
at the wavelengths covered by W1.

\clearpage

\subsection{D-type asteroids}

Asteroids with D-type taxonomic classifications become increasingly
common as distance from the Sun grows, from the Main Belt through the
Jupiter Trojan population \citep{demeo13}.  These objects, especially
the Jupiter Trojans, were likely implanted from a more distant
reservior during the early chaotic evolution of the Solar system
\citep{morbi05} and thus represent primitive material distinct from
objects that formed in the warmer region of the Main Belt.  NIR albedo
can be used to probe the distribution of these objects and
differentiate between classes of primitive bodies.  \citet{grav12tax}
compare $p_{W1}$ and $p_{W2}$ to distinguish asteroids with D-type
taxonomic classification from those with C- and P-type, and are able
to determine the overall population fraction of D-type objects in the
Jupiter Trojan and Hilda populations.  They find that the majority of
Jupiter Trojans are D-type at all sizes, while the Hilda population
transitions from a minority of D-types at diameters $D>40~$km to a
majority at smaller sizes.

\begin{figure}[ht]
\begin{center}
\includegraphics[scale=0.5]{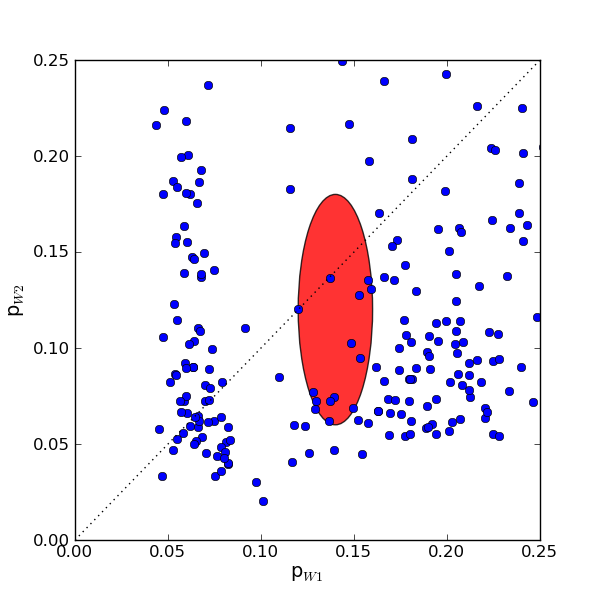}
\caption{W1 and W2 albedos for all measured objects.  The red ellipse
  marks the region populated by D-type asteroids as identified in
  \citet{grav12tax}.  This taxonomic classification shows no
  significant representation in the Main Belt objects studied here.}
\label{fig.TrojComp}
\end{center}
\end{figure}

Following \citet{grav12tax}, we show in Figure~\ref{fig.TrojComp} an
expanded view of the objects with lowest infrared albedos.  We
highlight the region of albedo-space that is occupied by D-type
asteroids in the Trojan and Hilda populations.  The diameter and
albedo fits from \citet{grav12tax} rely on the same model and
assumptions as we use here, and so comparisons between the two
populations should only depend on the random error associated with the
fits.  Only $2\%$ of all objects for which we measure $p_{W1}$ and
$p_{W2}$ fall in this region; with the exception of (114) Kassandra
and (267) Tirza (which are spectrally classified as T- and D-type
objects, respectively), all other candidate D-type objects are in the
outer Main Belt and have diameters between $10~$km$<D<40~$km,
consistent with the diameter regime where D-types dominate the Hilda
asteroids.  One object, (1755) Lorbach, is identified as an S-type in
\citet{neesePDS}, but this classification relies on only two optical
colors.  For the outer Main Belt, we do not see a significant
population of D-type objects like what is observed in the Hildas and
Trojans \citep{demeo13}, but this is expected from the lower
efficiency of dynamical implantation compared with the Hilda and
Trojan populations \citet{levison09}.

We find no objects in the inner Main Belt with albedos consistent with
D-type objects.  This is in contrast to the results of \citet{demeo14}
who find a small population of these bodies; however, this difference
can be understood through the selection effects in our survey.
Although our sample probes a large number of objects with semimajor
axis $a<2.5~$AU only a handful have low albedo.  Inner Main Belt low
albedo asteroids are more likely to have significant thermal emission
in W2, so we are not able to determine $p_{W2}$ for these objects and
they will not appear in our analysis.

\clearpage

\subsection{Low-$p_{W1}$/High-$p_{W2}$ objects}

Figure \ref{fig.TrojComp} shows a group of objects with low visible
and W1 albedos ($p_V,p_{W1}<0.1$) but high W2 albedo ($p_{W2}>0.1$),
which also appear as the objects with the reddest W1-W2 slopes in
Figure~\ref{fig.slopeComp}.  This class of object does not have an
analog in the Jupiter Trojan or Hilda populations \citep{grav12tax}
where we find parallels to other low-albedo MBA populations.  Objects
from this group that have spectroscopic or photometric taxonomic
classifications in PDS are typically designated as C-type or a related
subclass \citep{neesePDS}.  There are occasional objects with other
classifications such as X-, F- and S-, or dual classifications,
although often in these cases the designation is based on only 2 color
indices and so is of low reliability.  It is possible that this group
could represent a different class of objects that is not found in the
more distant Solar system populations, or instead could be a failure
of the thermal model to converge for certain objects with low albedos.

In order to test if these objects are a result of a failure of the
fitting routine, we take all objects in our fitted population with
$p_{W1}<0.1$ and compare the set with $p_{W2}\ge0.1$ to the set with
$p_{W2}<0.1$ (referred to as ``low-high'' and ``low-low''
respectively).  The low-high and low-low test sets are approximately
the same size (44 vs 48 objects), and have similar distributions of
semimajor axes, eccentricities, inclinations, $p_V$ and $p_{W1}$.  The
primary difference between these two groups is that the low-high
objects have significantly smaller heliocentric distances at the time
of observation than the low-low objects, resulting in higher subsolar
temperatures.  The diameters of the low-high objects are also
characteristically smaller than those of the low-low group, however we
cannot distinguish if this is an actual difference between the groups
or is a change in sensitivity as a result of the low-high objects
being closer to the Sun and telescope at the time of observation, and
thus warmer and brighter.

The asteroid (656) Beagle is a particularly interesting case for
testing the differences between these two sets of objects.  NEOWISE
observed this asteroid at two different epochs, both while fully
cryogenic, with good sensitivity at all four bands.  One epoch of
observations results in a NEATM best-fit that falls into the low-low
group, while the other epoch falls into the low-high group.  The
low-high epoch data were taken when Beagle was $0.21~$ AU closer to the
Sun ($2.82~$AU vs $3.03~$AU for the low-low case), following the trend
seen for the overall population.  The best-fit for NEATM in the
low-high epoch has a beaming parameter of $\eta=1.46$ and a diameter
of $D=62~$km while the low-low epoch has best-fit values of
$\eta=1.03$ and $D=48~$km which is the reverse of the diameter trend
mentioned above.  This large disagreement in diameter is not
unexpected given the difference in best-fit beaming parameter which is
inversely proportional to the fourth power of the subsolar temperature
used in the NEATM model, and thus will change the model's emitted
flux.

The observations used for our fits were visually inspected, as well as
compared to the WISE all-sky atlas of stationary sources, and show no
significant contamination by background stars or galaxies.  We note
that (656) Beagle has a large amplitude lightcurve (A$>1~$mag) and a
period of $7.035~$hours \citep{menke05}.  Although large amplitudes
can increase uncertainty in the fits, our data consist of 12 data
points over $1$ day and 15 data points over $1.25$ days, so both
epochs cover multiple rotations.  As such, light curve variations
should be averaged over by our fits, and should only contribute a
small amount to the total uncertainty in the fit.

As a test of our model, we perform a NEATM fit using only bands W1,
W2, and W3 as constraints, assuming the W4 measurements are
anomalously high, and a fixed beaming parameter of $\eta=1.0$.  When
using a fixed beaming parameter we cannot adequately constrain
$p_{W2}$, and so assume it is equal to $p_{W1}$.  For these restricted
fits, both epochs converge to diameters that agree to within $10\%$,
but they cannot reproduce the measured magnitudes as well as the
full-fit case.  As we are using one fewer constraint but two fewer
variables, this is not surprising.  The fits for (656) Beagle given by
\citet{masiero11} are nearly identical to these restricted fits, but
also cannot fully reproduce the measured magnitudes, particularly for
the low-high epoch.  Restricting our model further and only fitting W1
and W3, we find that both epochs converge to nearly identical
diameters, and visible and infrared albedos.
 
We can understand these results by looking at where the best-fit model
deviates from the data.  For the low-high epoch, the full NEATM fit
cannot reproduce the W2, W3, and W4 fluxes simultaneously, with the W2
and W4 measurements showing excesses not observed in W3.  Our full
model finds a best fit solution allowing W3 and W4 to determine the
diameter and beaming which under-produces flux in W2, but corrects that
by increasing $p_{W2}$.  If we ignore the W4 measurements, the W2 and
W3 fluxes still cannot be reproduced in the low-high epoch solely with
thermal emission and reflected light without resorting to extreme
changes in $p_{W2}$.

One possible explanation for the disagreement between epochs is that
we are observing significant differences between the thermal emission
in the morning and afternoon hemispheres of the asteroid.  If (656)
Beagle has a relatively high thermal inertia, there may be a
significant lag to the thermal re-emission of incident light which is
not accounted for in the NEATM model.  Our two epochs of observation
are at phase angles of $\alpha\sim20^\circ$, but on opposite sides of
the body.  (656) Beagle is on a low-inclination orbit, so if we assume
the rotation pole is oriented perpendicular to the orbital and
ecliptic planes and that the rotation is prograde, then the data from the
low-high epoch would correspond to the afternoon hemisphere and the
data from the low-low epoch would correspond to the morning
hemisphere.  Future work will implement a full thermophysical model of
this object to test if the W2 and W4 excesses can be explained by a
morning/afternoon dichotomy.  For all other objects in the low-high
group which were only observed at a single epoch, we cannot currently
differentiate between poor fits to the beaming parameter and actual
excesses in the W2 and/or W4 bands.

An alternate possibility is that these fits are indicative of problems
with the flux measurement of partially saturated sources in the WISE
data.  \citet{cutriExpSupp} discuss the process by which fluxes are
measured for saturated sources through PSF-fitting photometry.  Flux
measurements are available for sources many magnitudes above the
brightness where the central pixel saturates through fitting of the
PSF wings, however for very bright sources in bands W2 and W3, there
appears to be a slight over-estimation of the fluxes.  None of the
objects we fit here had W2 magnitudes in this saturated regime,
however the majority of objects with $p_{W1}<0.1$ had W3 magnitudes in
this problematic region.  

We correct for saturation estimation issues in our thermal model,
however there is the potential that the error for asteroidal sources
cannot be adequately described by this correction, which was
calibrated for stars.  The difference in the spectral energy
distributions through the W3 bandpass of hot, blue stars and cooler,
red asteroids potentially could result in differences deep in the
wings of the PSF for each type of source that are not fully
encompassed by the color correction.  These subtle changes can have a
significant impact on saturated sources where only the wings are
available for profile-fitting, however there are an insufficient
number of well-calibrated, W3-bright sources with the appropriate
spectral energy distribution to correct for this effect.  Although
this error may only have a small effect on other physical parameters
within our modeled systematic uncertainties, due to W2's position on
the Wien's side of Main Belt asteroid thermal emission for some of our
objects, a small change in W3 can result in a large change in W2 flux,
and thus our interpretation of the W2 albedo.  As such, caution is
strongly encouraged in interpreting fits for objects with very bright
W3 magnitudes (W3$<4~$mag).

\subsection{NIR Albedos of Asteroid Families}
\label{sec.families}

The distributions of visible albedos for members of each asteroid
family have much narrower spread than the albedo distribution of the
Main Belt as a whole \citep{masiero11} as is expected from a
population resulting from the collisional breakup of a single parent
body.  As the (4) Vesta family shows a narrow albedo distribution but
originated from a differentiated body, we do not expect the albedo
distributions of other cratering-event families that may have been
partially- or fully-melted to differ significantly from
non-differentiated families.  It is possible that families formed from
the complete disruption of a differentiated parent body may show a
broader albedo distribution, though we do not see any evidence for a
case like this in our data.  Visible albedo can also be used to
improve family membership lists by rejecting outlier objects that are
dynamically similar to the family \citep{masiero13,walsh13}.  Using
the refined family lists from \citet{masiero13} we investigate the
distribution of $p_{W1}$ for families as a more accurate tracer of the
surface properties of these asteroids.

\begin{figure}[ht]
\begin{center}
\includegraphics[scale=0.5]{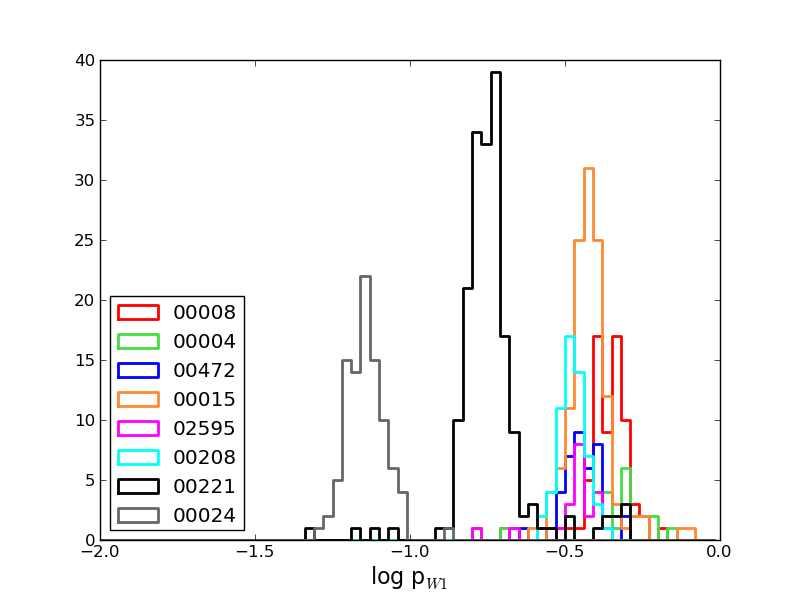}
\caption{W1 albedo distributions for $8$ asteroid families with more
  than $20$ measured NIR albedos, identified by the family ID given in
  \citet{masiero13} and Table~\ref{tab.famprop}.  Families show narrow
  distributions of albedos correlating with one of three major albedo
  groupings.}
\label{fig.nirFamHist}
\end{center}
\end{figure}

Figure~\ref{fig.nirFamHist} shows the distribution of $p_{W1}$ albedos
for the $8$ families where more than $20$ members had a $p_{W1}$
albedo measurement.  Asteroid families break into three clear
groupings, following the three peaks in the albedo distribution shown
in Figure~\ref{fig.scatAlb}.  Our dataset depends on reflected light
measurements, so high-albedo families are over-represented in the
distribution compared with the population of all known families, which is dominated by low-albedo families.  The only low
NIR-albedo family with more than $20$ measured objects was (24)
Themis, however other families such as (10) Hygiea, (145) Adeona,
(276) Adelheid, (511) Davida, (554) Peraga (equivalent to other lists'
Polana family), and (1306) Scythia also show low NIR albedos, but
these families contain only a small number of objects with measured
$p_{W1}$.  The families (4) Vesta, (8) Flora, (15) Eunomia, (208)
Lacrimosa, (472) Roma, and (2595) Gudiachvili all have high $p_{W1}$
and show only a small spread in mean albedo, while (135) Hertha
(equivalent to other lists' Nysa family) and (254) Augusta join them
at a lower significance level.

The (221) Eos family is the only one of the large families to have a
moderate NIR albedo, in between the high- and low-albedo populations,
indicating that this family has surface properties that are rare among
the large Main Belt asteroids.  The $p_{W1}$ values for this family
confirm the observed moderate visible albedo as a separate grouping
that could not be conclusively distinguished from the high $p_V$
population by \citet{masiero13}.  The Eos family parent has a K-type
spectral taxonomy in the Bus-DeMeo system \citep{demeo09}.  K-type
objects are considered `end-members' of the classification scheme, and
have a $1~\mu$m absorption feature typically associated with silicates
such as olivine, but are distinct in spectroscopic principal component
space from the majority of S-class objects.  \citet{clark09} and
\citet{hardersen11} associate K-type objects with the parent body of
carbonaceous chondrite meteorites, specifically CO chondrites, while
\citet{mothediniz05} show evidence that (221) Eos may have been
partially differentiated.  \citet{broz13eos} calculate the time since
the breakup of the (221) Eos family as $1.5-1.9~$Gyr, making it one of
the oldest Main Belt families with a measured age.  These observed
properties, when taken together, paint the Eos family as having a
unique evolutionary history that can be studied using remote
observations in combination with hand samples from the meteorite
record to trace the early history of the Solar system.

\clearpage

We note that approximately half of the objects fit for the (298)
Baptistina family had albedos similar to the Eos family, while the
remainder appear to be drawn from the high-albedo group.  This result
is based on only a small number of measured Baptistina members, and
thus is not conclusive, however if confirmed would further impede
attempts to assign a unique composition to this family
\citep[cf.][]{reddy11baf} or determine its age and evolution
\citep[cf.][]{masiero12bap}.

\begin{figure}[ht]
\begin{center}
\includegraphics[scale=0.55]{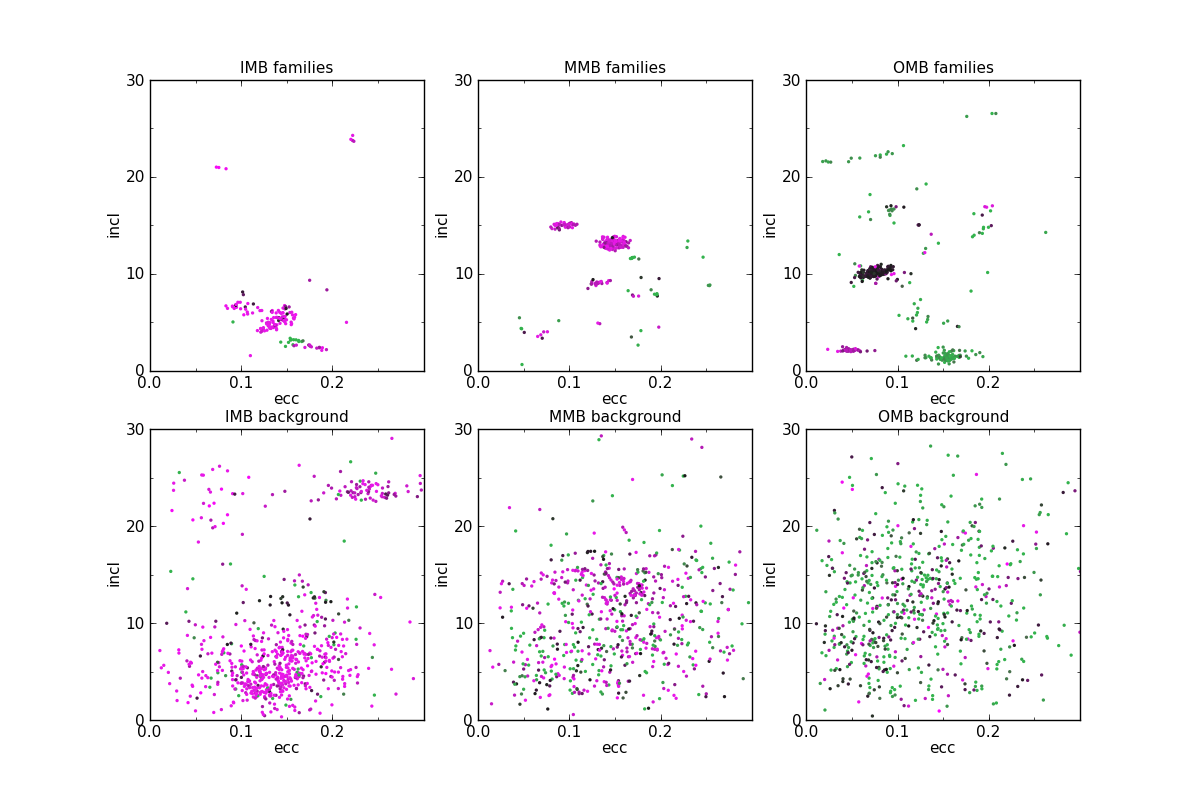}
\caption{Proper orbital inclination (incl) vs eccentricity (ecc) for
  inner- (left), middle- (center), and outer-Main Belt populations
  (right), for objects associated with families (top) and background
  objects not linked to families (bottom).  Colors of the points map
  the W1 albedo (from green to black to magenta for increasing
  $p_{W1}$), following Figure~\ref{fig.w1alb}a.}
\label{fig.eiFamBack}
\end{center}
\end{figure}

Figure~\ref{fig.eiFamBack} shows the proper orbital eccentricity and
inclination of all objects with measured $p_{W1}$.  The Main Belt is
split into three regions by proper semi-major axis ($a$): the
inner-Main Belt (IMB, $1.8~$AU$<a<2.5~$AU), the middle-Main Belt (MMB,
$2.5~$AU$<a<2.82~$AU), and the outer-Main Belt (OMB,
$2.82~$AU$<a<3.6~$AU).  We show separately the objects that were
associated with an asteroid family by \citet{masiero13} and those that
are members of the background population.  The (221) Eos family stands
out distinctly in the belt, although objects with similar $p_{W1}$ are
present in the background population in all three regions.

These plots show the clear trend of albedo decreasing with distance
from the Sun, however our observational bias against small, low albedo
objects amplifies this effect.  \citet{broz13fams}, \citet{carruba13},
and \citet{masiero13} observe halos of objects beyond the limits of
typical family-identification techniques, however we do not see
evidence for these halos in the background population in our dataset.
Halos are typically associated with asteroids that have dispersed a
large distance from the family center via Yarkovsky and gravitational
forces, which will have smaller diameters than objects that were above
our sensitivity limit for $p_{W1}$ determination.  Our significantly
smaller sample size than what is typically used in surveys
investigating family halos may also contribute to their absence in our
data.

\clearpage

In Table~\ref{tab.famprop} we present the orbital and physical
properties for all families identified in \citet{masiero13} that had
at least one member with a fitted NIR albedo.  We list the name of the
family, average proper orbital elements, largest (D$_{max}$) and
smallest diameter (D$_{min}$) represented in our sample, W1 albedos
with standard deviations, and number of family members with data
sufficient to fit.  We also provide for reference the mean $p_V$ and
standard deviation from \citet{masiero13}.  For cases where only a
single body had a measured $p_{W1}$ (often but not always the parent
body of the family), D$_{min}$ is marked with a `...' entry and no
standard deviation is given for the mean W1 albedo for families with
less than 10 members.  Asteroids that have been incorrectly associated
with families may have very different mineralogies and thus spectral
behavior in the NIR, which could make those objects more likely to
fulfill the selection requirements for measured $p_{W1}$.  Thus,
particular caution is necessary when dealing with families suffering
small number statistics, especially families with only a single
$p_{W1}$-fit object.  We note that the mean $p_V$ albedos presented in
\citet{masiero13} are based on larger numbers of objects and so will
generally be more accurate than the mean $p_{W1}$ values given here.

It is also possible to use the W1 albedo to further refine family
memberships, particularly for confused cases such as the Nysa-Polana
complex.  \cite{masiero13} divided this complex into a high albedo
component with largest body (135) Hertha and a low albedo component
with largest body (554) Peraga which is nearly twice the diameter of
(142) Polana.  We use NIR albedo to reject objects from the low-albedo
family that had moderate visible albedos but W1 albedos characteristic
of the high-albedo family.  Asteroids (261), (1823), (2717), and
(15112) can thus be rejected as members of the (554) Peraga group
based on W1 albedo.  We note that because of a typo in
\citet{masiero13} (135) Hertha was mistakenly listed as associated
with (554) Peraga instead of with its own family, which we correct
here.  \citet{walsh13} present dynamical arguments to divide the (554)
Peraga family into two sub-families, however we are unable to see any
distinction between these groups in visible or W1 albedo.

\begin{table}[ht]
\begin{center}
\caption{Average orbital and physical properties for asteroid family
  members with measured $p_{W1}$.  Mean $p_V$ values are taken from
  \citet{masiero13}}
\vspace{1ex}
{\tiny
\noindent
\begin{tabular}{ccccccccc}
\tableline
Family & semimajor axis & eccentricity & inclination & D$_{max}$ & D$_{min}$& $<p_V>$ & $<p_{W1}>$ & Sample\\
 &  (AU) &  &  (deg) & (km)& (km)&  &  &  Size\\
\tableline
00434 & 1.937 & 0.077 & 20.947 &   7.62 &   3.01 &  0.725 $\pm$ 0.172 & 0.736         &   3\\
00254 & 2.197 & 0.122 &  4.202 &  11.85 &   5.71 &  0.298 $\pm$ 0.105 & 0.420         &   7\\
00008 & 2.244 & 0.141 &  5.251 & 155.74 &   3.86 &  0.291 $\pm$ 0.091 & 0.435 $\pm$ 0.074 &  70\\
00298 & 2.256 & 0.148 &  5.723 &  12.32 &   5.36 &  0.146 $\pm$ 0.034 & 0.284         &   9\\
00163 & 2.325 & 0.215 &  5.008 &   4.78 & ... &  0.053 $\pm$ 0.016 & 0.401         &   1\\
00587 & 2.338 & 0.222 & 23.901 &  12.23 &   4.41 &  0.310 $\pm$ 0.090 & 0.421         &   4\\
01646 & 2.353 & 0.102 &  8.002 &  12.47 &  11.57 &  0.204 $\pm$ 0.081 & 0.221         &   2\\
00004 & 2.366 & 0.101 &  6.518 & 109.51 &   4.53 &  0.357 $\pm$ 0.110 & 0.466 $\pm$ 0.129 &  22\\
00012 & 2.390 & 0.185 &  8.853 & 126.64 &   7.62 &  0.066 $\pm$ 0.021 & 0.318         &   2\\
00135 & 2.405 & 0.178 &  2.455 &  82.15 &   3.77 &  0.280 $\pm$ 0.088 & 0.372 $\pm$ 0.071 &  13\\
00302 & 2.407 & 0.110 &  1.576 &   6.11 & ... &  0.062 $\pm$ 0.021 & 0.422         &   1\\
00554 & 2.418 & 0.157 &  3.049 &  55.18 &   11.75 &  0.061 $\pm$ 0.021 & 0.054 $\pm$ 0.012 &  13\\
00752 & 2.463 & 0.091 &  5.049 &  60.85 & ... &  0.053 $\pm$ 0.014 & 0.048         &   1\\
13698 & 2.469 & 0.118 &  6.534 &   5.31 & ... &  0.367 $\pm$ 0.098 & 0.412         &   1\\
01658 & 2.560 & 0.172 &  7.749 &  13.81 &   5.97 &  0.255 $\pm$ 0.074 & 0.343         &   3\\
00472 & 2.562 & 0.094 & 15.009 &  47.04 &   6.25 &  0.261 $\pm$ 0.079 & 0.363 $\pm$ 0.051 &  39\\
00005 & 2.576 & 0.198 &  4.514 & 113.00 & ... &  0.240 $\pm$ 0.105 & 0.365         &   1\\
00606 & 2.587 & 0.179 &  9.631 &  39.53 & ... &  0.117 $\pm$ 0.028 & 0.137         &   1\\
05079 & 2.601 & 0.247 & 11.730 &  14.76 & ... &  0.068 $\pm$ 0.020 & 0.047         &   1\\
00404 & 2.628 & 0.229 & 13.062 & 105.41 &  73.07 &  0.060 $\pm$ 0.025 & 0.059         &   2\\
00015 & 2.630 & 0.149 & 13.181 & 299.21 &   4.45 &  0.263 $\pm$ 0.084 & 0.382 $\pm$ 0.073 & 126\\
00569 & 2.634 & 0.175 &  2.659 &  13.18 & ... &  0.054 $\pm$ 0.016 & 0.064         &   1\\
00145 & 2.676 & 0.170 & 11.642 & 132.59 &  14.70 &  0.062 $\pm$ 0.018 & 0.055         &   8\\
00410 & 2.727 & 0.253 &  8.824 &  27.28 &   8.49 &  0.085 $\pm$ 0.028 & 0.075         &   3\\
00539 & 2.739 & 0.164 &  8.274 &  56.04 & ... &  0.061 $\pm$ 0.023 & 0.039         &   1\\
00396 & 2.742 & 0.168 &  3.497 &  37.29 & ... &  0.093 $\pm$ 0.024 & 0.115         &   1\\
00808 & 2.744 & 0.132 &  4.902 &  37.68 &   9.51 &  0.232 $\pm$ 0.071 & 0.380         &   2\\
00363 & 2.750 & 0.045 &  5.480 &  19.34 & ... &  0.068 $\pm$ 0.018 & 0.079         &   1\\
00128 & 2.750 & 0.088 &  5.181 & 193.08 & ... &  0.075 $\pm$ 0.024 & 0.071         &   1\\
01734 & 2.769 & 0.194 &  7.951 &  23.82 &   8.72 &  0.056 $\pm$ 0.017 & 0.074         &   6\\
00847 & 2.777 & 0.070 &  3.742 &  30.08 &   8.78 &  0.218 $\pm$ 0.075 & 0.359         &   5\\
00272 & 2.783 & 0.048 &  4.232 &  25.67 &  21.09 &  0.047 $\pm$ 0.014 & 0.109         &   3\\
00322 & 2.783 & 0.198 &  9.521 &  73.15 & ... &  0.078 $\pm$ 0.027 & 0.193         &   1\\
01128 & 2.788 & 0.048 &  0.659 &  48.63 & ... &  0.048 $\pm$ 0.013 & 0.046         &   1\\
02595 & 2.791 & 0.132 &  9.068 &  14.62 &   6.59 &  0.262 $\pm$ 0.075 & 0.343 $\pm$ 0.059 &  21\\
01668 & 2.806 & 0.178 &  4.152 &  25.83 & ... &  0.052 $\pm$ 0.014 & 0.048         &   1\\
03985 & 2.851 & 0.123 & 15.052 &  22.11 &  16.19 &  0.176 $\pm$ 0.058 & 0.213         &   3\\
00081 & 2.854 & 0.180 &  8.233 & 123.96 & ... &  0.056 $\pm$ 0.016 & 0.053         &   1\\
00208 & 2.875 & 0.048 &  2.129 &  49.99 &  10.36 &  0.237 $\pm$ 0.063 & 0.335 $\pm$ 0.032 &  59\\
00845 & 2.940 & 0.036 & 11.999 &  58.53 & ... &  0.061 $\pm$ 0.017 & 0.055         &   1\\
00179 & 2.972 & 0.076 &  9.027 &  74.58 & ... &  0.223 $\pm$ 0.069 & 0.326         &   1\\
00816 & 3.004 & 0.145 & 13.168 &  50.09 & ... &  0.051 $\pm$ 0.026 & 0.054         &   1\\
00221 & 3.020 & 0.077 & 10.181 &  95.62 &  10.03 &  0.158 $\pm$ 0.048 & 0.190 $\pm$ 0.065 & 186\\
00283 & 3.070 & 0.109 &  8.917 & 145.55 &  17.68 &  0.048 $\pm$ 0.019 & 0.084         &   2\\
02621 & 3.086 & 0.128 & 12.128 &  47.92 & ... &  0.081 $\pm$ 0.029 & 0.066         &   1\\
01113 & 3.112 & 0.137 & 14.093 &  48.37 & ... &  0.074 $\pm$ 0.031 & 0.350         &   1\\
00780 & 3.117 & 0.070 & 18.186 & 114.26 & ... &  0.056 $\pm$ 0.018 & 0.060         &   1\\
01040 & 3.122 & 0.197 & 16.728 &  22.67 &   7.56 &  0.225 $\pm$ 0.075 & 0.348         &   4\\
00511 & 3.138 & 0.192 & 14.439 & 285.84 &  23.46 &  0.065 $\pm$ 0.026 & 0.076         &   8\\
00024 & 3.142 & 0.153 &  1.457 & 193.54 &  17.43 &  0.068 $\pm$ 0.021 & 0.073 $\pm$ 0.012 &  95\\
00928 & 3.143 & 0.193 & 16.359 &  62.54 &  24.29 &  0.075 $\pm$ 0.038 & 0.057         &   2\\
00010 & 3.143 & 0.130 &  5.514 & 153.58 &  16.69 &  0.070 $\pm$ 0.023 & 0.079 $\pm$ 0.035 &  17\\
03330 & 3.154 & 0.199 & 10.149 &  15.49 & ... &  0.044 $\pm$ 0.015 & 0.053         &   1\\
00490 & 3.165 & 0.061 &  9.323 &  79.87 & ... &  0.069 $\pm$ 0.022 & 0.050         &   1\\
00778 & 3.169 & 0.262 & 14.282 &  19.36 & ... &  0.066 $\pm$ 0.020 & 0.070         &   1\\
01306 & 3.170 & 0.091 & 16.448 &  72.24 &  15.54 &  0.061 $\pm$ 0.021 & 0.107 $\pm$ 0.056 &  14\\
00031 & 3.177 & 0.196 & 26.445 & 281.98 &  17.31 &  0.057 $\pm$ 0.016 & 0.068         &   3\\
00618 & 3.189 & 0.058 & 15.879 & 131.23 & ... &  0.056 $\pm$ 0.018 & 0.063         &   1\\
00702 & 3.190 & 0.021 & 21.598 & 196.47 &  26.15 &  0.066 $\pm$ 0.022 & 0.071         &   3\\
00276 & 3.190 & 0.072 & 22.189 & 100.36 &  21.98 &  0.068 $\pm$ 0.022 & 0.073 $\pm$ 0.009 &  11\\
01303 & 3.215 & 0.126 & 19.023 & 102.43 &  28.58 &  0.049 $\pm$ 0.017 & 0.069         &   2\\
00087 & 3.485 & 0.054 &  9.846 & 288.38 & ... &  0.057 $\pm$ 0.017 & 0.082         &   1\\
\hline
\end{tabular}
}
\label{tab.famprop}
\end{center}
\end{table}

\section{Conclusions}

We present revised thermal model fits for Main Belt asteroids,
allowing for the albedo in each of the near-infrared reflected
wavelengths to be fit independently.  The $3.4~\mu$m and $4.6~\mu$m
spectral regions covered by the WISE/NEOWISE W1 and W2 bandpasses are
poorly probed in ground-based spectroscopy but can be used to provide
insight into asteroid mineralogical composition by constraining
spectral slope.  In total we present $3080$ fits of $p_{W1}$ and/or
$p_{W2}$ for $2835$ unique Main Belt objects.  

The MBA population has three distinct peaks in our observed $p_{W1}$
distribution at $p_{W1}\sim0.06$, $p_{W1}\sim0.16$, and
$p_{W1}\sim0.4$.  The high and low $p_{W1}$ peaks correspond to the
high and low visible albedo groups observed previously, while the
moderate $p_{W1}$ peak corresponds to an intermediate visible albedo
that is blended with the high $p_V$ objects in visible albedo
distributions.  The distribution of albedos we measure have a larger
fraction of high-albedo objects than what was observed for the MBA
visible albedo distribution, however this is an effect of the biases
in our sample selection.

Asteroid families have narrow $p_{W1}$ distributions corresponding to
one of the three observed $p_{W1}$ peaks.  The (221) Eos family
represents the only significant concentration of objects near the peak
at $p_{W1}\sim0.16$, although other objects with this albedo that are
not related to asteroid families are scattered throughout the entire
Main Belt region.  This family also corresponds to an unusual `end
member' taxonomic classification, K-type, that has been suggested to
correspond to a partially differentiated parent or olivine-rich
mineralogy.  NIR albedo measurements provide a way to rapidly search
the known population for candidate K-type objects in the Main Belt,
and are a powerful tool that acts as a proxy for asteroid taxonomic
type.

Our results show that the majority of high albedo objects, believed to
have surface compositions dominated by silicates and similar to
ordinary chondrite meteorites, show an overall reddening from visible
to W1 wavelengths similar to what is seen in the NIR.  The spectra
become blue from W1 to W2, which is also seen in some meteorite
populations, particularly the Eucrites.  This overall picture is
consistent with a primarily-silicate dominated composition. Objects
with moderate infrared albedos show similar behavior across the
wavelengths probed here, although the lower albedo value at W1 may
indicate subtle differences in composition from the high albedo
population or even a mix of different mineralogies.

The low albedo objects in our sample show a much wider range of
behavior in these spectral regions.  Many object show red slopes
across all wavelengths consistent with the NIR spectral behavior of
C/D/P-type objects.  However approximately $10\%$ of our population
show a blue slope from visible to W1, even in spite of the biases
against blue-sloped, low-albedo objects in our sample.  These objects
are associated with B and Ch spectral taxonomies.  The blue
visible-to-W1 spectral slope in the Ch class objects may be indicative
of a significant absorption feature at W1 wavelengths from minerals
such as carbonates.

The fits presented here are based on reflected light, and thus our
sample will not accurately represent the true distribution of $p_{W1}$
or $p_{W2}$.  Small, low albedo asteroids as well as objects with blue
NIR spectral slopes are more likely to be undetected in the W1 and/or
W2 wavelengths and thus underrepresented in our population
distributions.  A larger survey with greater sensitivity in these
spectral regions is required to extend these results to a population
comparable to the one with measured diameters and visible albedos.

\section*{Acknowledgments}
JM was partially supported by a NASA Planetary Geology and Geophysics
grant.  CN, RS, and SS were supported by an appointment to the NASA
Postdoctoral Program at JPL, administered by Oak Ridge Associated
Universities through a contract with NASA.  We thank the referee for
the helpful comments that greatly improved this manuscript.  This
publication makes use of data products from the Wide-field Infrared
Survey Explorer, which is a joint project of the University of
California, Los Angeles, and the Jet Propulsion Laboratory/California
Institute of Technology, funded by the National Aeronautics and Space
Administration.  This publication also makes use of data products from
NEOWISE, which is a project of the Jet Propulsion
Laboratory/California Institute of Technology, funded by the Planetary
Science Division of the National Aeronautics and Space Administration.
This research has made use of the NASA/IPAC Infrared Science Archive,
which is operated by the Jet Propulsion Laboratory, California
Institute of Technology, under contract with the National Aeronautics
and Space Administration.



\clearpage
\begin{thebibliography}{XXX}

\bibitem[Bowell \etal(1989)]{bowell89}
Bowell, E., Hapke, B., Domingue, D., Lumme, K., Peltoniemi, J. \& Harris, A.W., 
1989, Asteroids II, University of Arizona Press, 524.

\bibitem[Broz \& Morbidelli(2013a)]{broz13eos}
Broz, M. \& Morbidelli, A., 2013a, Icarus, 223, 844

\bibitem[Broz \etal(2013b)]{broz13fams}
Broz, M., Morbidelli, A., Bottke, W.F., Rozehnal, J., Vokrouhlick\'{y} \& D., Nesvorn\'{y}, D., 2013b, A\&A, 551, A117

\bibitem[Campins \etal(2010)]{campins10}
Campins, H., Hargrove, K., Pinilla-Alonso, N., Howell, E.S., \etal, 2010, Nature, 464, 1320.

\bibitem[Carruba \etal(2013)]{carruba13}
Carruba, V., Domingos, R.C., Nesvorny, D., Roig, F., Huaman, M.E., Souami, D., 2013, MNRAS, 433, 2075.

\bibitem[Clark \etal(2009)]{clark09}
Clark, B.E., Ockert-Bell, M.E., Cloutis, E.A., \etal, 2009, Icarus, 202, 119.

\bibitem[Cutri \etal(2012)]{cutriExpSupp}
Cutri, R.M., Wright, E.L., Conrow, T., Bauer, J., \etal, 2012, ``Explanatory Supplement to the WISE All-Sky Data Release Products'', {\it http://wise2.ipac.caltech.edu/docs/release/allsky/expsup/index.html}.

\bibitem[DeMeo \etal(2009)]{demeo09}
DeMeo, F.E., Binzel, R.P., Slivan, S.M. \& Bus, S.J., 2009, Icarus, 202, 160.

\bibitem[DeMeo \& Carry(2013)]{demeo13}
DeMeo, F.E. \& Carry, B., 2013, Icarus, 226, 723.

\bibitem[DeMeo \etal(2014)]{demeo14}
DeMeo, F.E., Binzel, R.P., Carry, B., Polishook, D., \& Moskovitz, N.A, 2014, Icarus, 229, 392

\bibitem[Gaffey \etal(2002)]{gaffey02}
Gaffey, M.J., Cloutis, E.A., Kelley, M.S. \& Reed, K.L., 2002, Asteroids III, W. F. Bottke Jr., A. Cellino, P. Paolicchi, and R. P. Binzel (eds), University of Arizona Press, 183.

\bibitem[Grav \etal(2011a)]{grav11troj}
Grav, T., Mainzer, A., Bauer, J., \etal, 2011a, ApJ, 742, 40.

\bibitem[Grav \etal(2011b)]{grav11hilda}
Grav, T., Mainzer, A., Bauer, J., \etal, 2011b, ApJ, 744, 197.

\bibitem[Grav \etal(2012)]{grav12tax}
Grav, T., Mainzer, A.K., Bauer, J.M., Masiero, J.R., Nugent, C.R., 2012, ApJ, 759, 49.

\bibitem[Hardersen \etal(2011)]{hardersen11}
Hardersen, P.S, Cloutis, E.A., Reddy, V., \etal, 2011, M\&PS, 46, 1910.

\bibitem[Harris(1998)]{harrisNEATM}
Harris, A.W., 1998, Icarus, 131, 291.

\bibitem[Harris \& Drube(2014)]{harris14}
Harris, A.W. \& Drube, L., 2014, ApJL, 785, 4

\bibitem[Levison \etal(2009)]{levison09}
Levison, H.F., Bottke, W.F., Gounelle, M., \etal, 2009, Nature, 460, 364.

\bibitem[Mainzer \etal(2011a)]{mainzer11nw}
Mainzer, A.K., Bauer, J.M., Grav, T., Masiero, J., Cutri, R.M., Dailey, J., Eisenhardt, P., McMillan, R.M. \etal, 2011a, ApJ, 731, 53.

\bibitem[Mainzer \etal(2011b)]{mainzer11cal}
Mainzer, A.K., Grav, T., Masiero, J., Bauer, J.M., Wright, E., Cutri, R.M., McMillan, R.S., Cohen, M., Ressler, M., Eisenhardt, P., 2011b, ApJ, 736, 100.

\bibitem[Mainzer \etal(2011c)]{mainzer11iras}
Mainzer, A.K., Grav, T., Masiero, J., Bauer, J.M., Wright, E., Cutri, R.M., Walker, R. \& McMillan, R.S., 2011c, ApJL, 737, 9.

\bibitem[Mainzer \etal(2011d)]{mainzer11tax}
Mainzer, A.K., Grav, T., Masiero, J., \etal, 2011d, ApJ, 741, 90.

\bibitem[Mainzer \etal(2011e)]{mainzer11neo}
Mainzer, A., Grav, T., Bauer, J.M., Masiero, J., \etal, 2011e, ApJ, 743, 156.

\bibitem[Mainzer \etal(2012)]{mainzer12pc}
Mainzer, A.K., Grav, T., Masiero, J., Bauer, J.M., Cutri, R.M., McMillan, R.S., Nugent, C., Tholen, D., Walker, R. \& Wright, E.L., 2012, ApJL, 760, 12.

\bibitem[Mainzer \etal(2014)]{mainzer14restart}
Mainzer, A.K., Bauer, J.M., Grav, T., Masiero, J., \etal, 2014, ApJ, submitted.

\bibitem[Masiero \etal(2011)]{masiero11}
Masiero, J.R., Mainzer, A.K., Grav, T., Bauer, J.M., Cutri, R.M., Dailey, J., Eisenhardt, P.R.M, \etal, 2011, ApJ, 741, 68.

\bibitem[Masiero \etal(2012a)]{masiero12pc}
Masiero, J.R., Mainzer, A.K., Grav, T., Bauer, J.M., Cutri, R., Nugent, C. \& Cabrera, M.S., 2012, ApJ, 759, L8.

\bibitem[Masiero \etal(2012b)]{masiero12bap}
Masiero, J.R., Mainzer, A.K., Grav, T., Bauer, J.M. \& Jedicke, R., 2012, ApJ, 759, 14.

\bibitem[Masiero \etal(2013)]{masiero13}
Masiero, J.R., Mainzer, A.K., Bauer, J.M., Grav, T., Nugent, C., Stevenson, R., 2013, ApJ, 770, 7.

\bibitem[Menke(2005)]{menke05}
Menke, J., 2005, Minor Planet Bulletin, 32, 85.

\bibitem[Morbidelli \etal(2005)]{morbi05}
Morbidelli, A., Levison, H.F., Tsiganis, K. \& Gomes, R., 2005, Nature, 435, 462.

\bibitem[Moth\'{e}-Diniz(2005)]{mothediniz05}
Moth\'{e}-Diniz, T. \& Carvano, J.M., 2005, A\&A, 442, 727.

\bibitem[Neese(2010)]{neesePDS}
Neese, C., Ed., Asteroid Taxonomy V6.0. EAR-A-5-DDR-TAXONOMY-V6.0. NASA Planetary Data System, 2010.

\bibitem[Reddy \etal(2011)]{reddy11baf}
Reddy, V., Carvano, J.M., Lazzaro, D., \etal, 2011, Icarus, 216, 184.

\bibitem[Reddy \etal(2012a)]{reddy12vesta}
Reddy, V., Nathues, A., Le Corre, L., \etal, 2012a, Science, 336, 700.

\bibitem[Reddy \etal(2012b)]{reddy12hed}
Reddy, V., Sanchez, J., Nathues, A., \etal, 2012b, Icarus, 217, 153.

\bibitem[Rivkin \& Emery(2010)]{rivkin10}
Rivkin, A.S. \& Emery, J.P., 2010, Nature, 464, 1322.

\bibitem[Walsh \etal(2013)]{walsh13}
Walsh, K.J., Delb\'{o}, M., Bottke, W.F., Vokrouhlick\'{y}, D. \& Lauretta, D.S., 2013, Icarus, 225, 283.

\bibitem[Wright \etal(2010)]{wright10}
Wright, E.L., Eisenhardt, P., Mainzer, A.K., Ressler, M.E., Cutri, R.M., Jarrett, T., Kirkpatrick, J.D., Padgett, D., \etal, 2010, AJ, 140, 1868.

\end{thebibliography}
\end{document}